\documentclass{article}[11pt]
\usepackage{graphicx} % Required for inserting images
\usepackage{geometry}[margin=0.4]
\usepackage{amsmath}
\usepackage{amssymb}
\usepackage{authblk}

\title{Towards model-based design of  causal manipulations of brain circuits with high spatiotemporal precision}
\author[1,*]{Anandita De}
\author[2,3]{Roozbeh Kiani}
\author[1,4,5]{Luca Mazzucato}

\affil[1]{Institute of Neuroscience, University of Oregon, Eugene, OR}
\affil[2]{Center for Neural Science, New York University, New York, NY}
\affil[4]{Depts. of Biology, Mathematics and Physics, University of Oregon, Eugene, OR}
\affil[5]{Physics Department and Padova Neuroscience Center, University of Padua, Italy}
\affil[*]{Corresponding author}

\date{}

\begin{document}

\maketitle
\begin{abstract}
Recent advancements in neurotechnology enable precise spatiotemporal patterns of microstimulations with single-cell resolution. The choice of perturbation sites must satisfy two key criteria: efficacy in evoking significant responses and selectivity for the desired target effects. This choice is currently based on laborious trial-and-error procedures, unfeasible for sequences of multi-site stimulations. Efficient methods to design complex perturbation patterns are urgently needed. Can we design a spatiotemporal pattern of stimulation to steer neural activity and behavior towards a desired target? We outline a method for achieving this goal in two steps. First, we identify the most effective perturbation sites, or hubs, only based on short observations of spontaneous neural activity. Second, we provide an efficient method to design multi-site stimulation patterns by combining approaches from nonlinear dynamical systems, control theory and data-driven methods. We demonstrate the feasibility of our approach using multi-site stimulation patterns in recurrent network models.
\end{abstract}

\section{Introduction}
Electrical stimulation is a cornerstone of therapeutic interventions for psychiatric and neurological disorders. Deep brain stimulation (DBS), for example, has demonstrated long-term efficacy in alleviating motor symptoms in Parkinson's disease by modulating dysfunctional neural circuits \cite{oehrn2024chronic}. Similarly, cortical microstimulation has advanced both basic and clinical neuroscience. In fundamental research, microstimulation has been pivotal in establishing causal links between neural activity and perceptual or cognitive processes---such as demonstrating that stimulating direction-selective neurons in the middle temporal (MT) area of macaque monkeys biases motion perception~\cite{salzman1990cortical,salzman1992microstimulation,murasugi1993microstimulation}. Clinically, microstimulation has been used to prevent epileptic seizures by restoring the balance between excitatory and inhibitory activity \cite{fisher2014electrical}. Recent advances in neurotechnology now enable the delivery of precise spatiotemporal stimulation patterns to control neural circuits, targeting specific neuronal ensembles—or even individual neurons—with remarkable accuracy. This precision, however, raises a fundamental question: how should we determine the optimal stimulation sites and patterns?

Effective stimulation sites must satisfy two key criteria: \textbf{selectivity} and \textbf{efficacy}. To study the causal impact of electrical stimulation on behavior, the targeted neurons must play a role in driving the specific neuronal or behavioral feature of interest. In sensory areas, neurons often exhibit selectivity for particular features of the sensory input and respond maximally at particular values of the stimulus feature. These tuning properties provide a principled guide for selecting stimulation sites to bias perception---for example, shifting motion perception by microstimulation of MT neurons \cite{salzman1990cortical,salzman1992microstimulation,murasugi1993microstimulation} or altering face perception by stimulating face-selective patches in inferotemporal cortex \cite{afraz2006microstimulation, parvizi2012electrical, moeller2017effect}.  

However, neurons in higher-order association areas often exhibit \textbf{mixed selectivity}, responding to combinations of sensory, motor, and cognitive variables \cite{rigotti2013importance,fusi2016neurons}. This complexity poses challenges for identifying suitable perturbation targets, as neurons may not exhibit straightforward, low-dimensional selectivity, or may require a large variety of behavioral tasks for accurate characterization of response selectivities.

In addition to selectivity, efficacy---the capacity of a stimulation site to influence network activity and downstream behavior---is critical. For instance, stimulating a \textbf{hub neuron}, defined by a large number of outgoing connections, can evoke widespread spiking activity and shift circuit dynamics more effectively than stimulating a sparsely connected neuron.

Traditionally, perturbation sites are identified through an iterative, trial-and-error process: stimulating one electrode at a time and monitoring its effects on neural activity and behavior. This approach is laborious, time-consuming, and may degrade electrode viability over time. A computational model capable of identifying candidate perturbation sites solely from recordings of spontaneous activity---without requiring complex behavioral tasks or prior stimulations---would significantly accelerate this process. Such a model would not only conserve experimental resources but also enable hypothesis-driven testing of whether hub neurons identified from resting-state activity indeed produce meaningful changes in ongoing network dynamics when perturbed. 

Below, we first present a systematic framework for identifying hub neurons or neural clusters (i.e., groups of nearby neurons with correlated activity) as promising targets for perturbation. We then combine nonlinear dynamical systems theory with linear control theory to design spatiotemporal stimulation patterns that steer network activity toward desired target states.

\subsection{Predicting high-efficacy sites from spontaneous activity}

One approach to identifying hub neurons is to infer causal connections between individual neurons or neural clusters using time series data. \textbf{Granger causality} (GC) has long been employed in neuroscience to infer directed interactions between brain regions \cite{brovelli2004beta}. However, GC assumes linear dependencies between time series, which may be insufficient to capture the full complexity of neural dynamics. To account for nonlinear interactions, \textbf{Transfer Entropy} (TE) was introduced as a model-free alternative for detecting directional influences \cite{schreiber2000measuring}. Both GC and TE are grounded in the assumption that neural activity---and thus the underlying time series---is stochastic in nature \cite{rolls2010noisy}.

\begin{figure}[h]
    \centering
    \includegraphics[width=\linewidth]{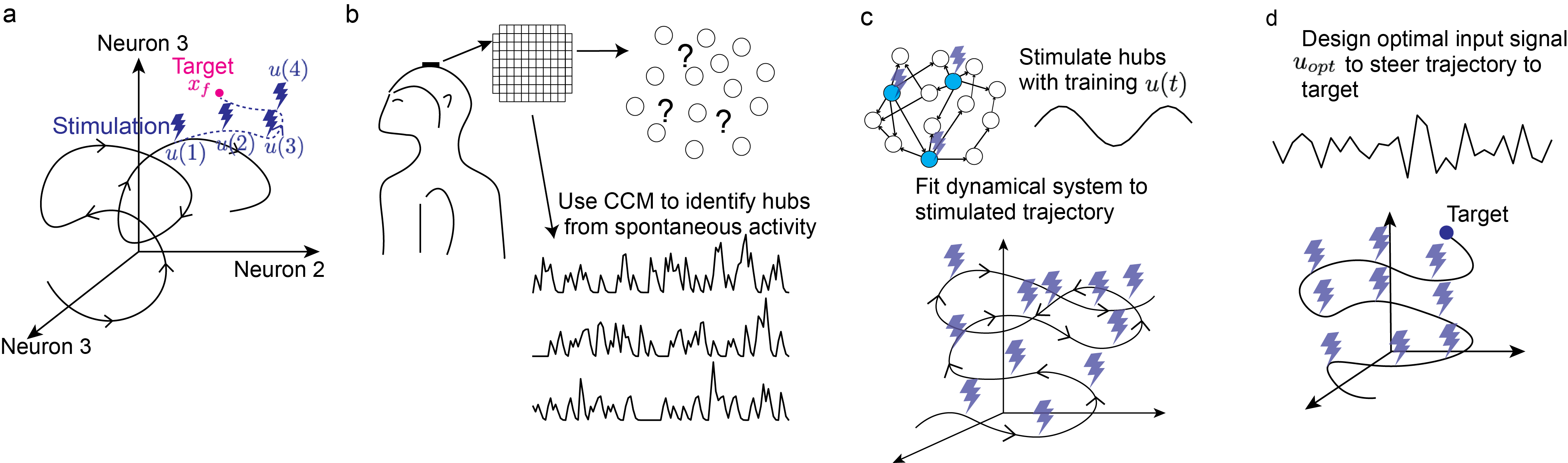}
    \caption{\textbf{a)} Schematic of a trajectory in neural state space. The goal is to design a stimulation sequence $u(1),\dots,u(T)$ that steers the trajectory to a desired target $x_f$. \textbf{b)} Schematic of an Utah array implanted into the cortex of a monkey. CCM identifies the hub electrodes from spontaneous activity recorded from this array. \textbf{c)} Schematic of hypothetical neural network that gives rise to the activity recorded by the Utah array. Training input signals $u(t)$ are injected into one of the hub nodes (top), evoking a trajectory in the neural state space (bottom). A dynamical system with inputs is fit to the data recorded from this training session, as discussed in Sections \ref{Koopman}, \ref{control}. \textbf{d)} The fitted dynamical system can be used to design an optimal control signal with minimum energy to steer the trajectory to a desired target. Schematic of optimal input $u_{opt}$ and the trajectory it evokes to reach the target state.}
    \label{fig:overview}
\end{figure}

Recent work has shown that neural activity during stimulation epochs can be directly modeled within an input-driven linear dynamical systems framework, with additional constraints based on cell-types \cite{o2022direct}. This approach provides insight into how stimulation influences population dynamics and, although it currently falls short of predicting single site efficacy solely from observations of spontaneous activity, it suggests a path toward that goal. Specifically, one could first identify the dynamical system underlying spontaneous activity in the absence of perturbations (e.g., using state space models or neural ODE \cite{kim2021inferring}). One could then model the stimulation as an external input driving the system's trajectory toward a target state to yield the desired behavioral effect (Figure \ref{fig:overview}a). \textbf{Control theory} \cite{liu2016control} offers a principled framework to design time-dependent stimulation patterns under constraints to achieve this objective. Recent studies have explored the possibility of whole-brain controllability within this framework \cite{gu2015controllability,manjunatha2024controlling}. However, these approaches have so far relied on the strong---an arguably unjustified---assumption of \textbf{linear neural dynamics}.

In contrast, an alternative view conceptualizes the brain as a \textbf{nonlinear dynamical system} with \textbf{chaotic dynamics} \cite{beggs2003neuronal,shew2009neuronal}. From a geometric perspective, neural activity can be represented as a trajectory through a neural state space, where each axis corresponds to a neuron and each point denotes the population activity at a given time. Under the chaotic dynamics hypothesis, this trajectory evolves on a \textbf{chaotic attractor}---a manifold with a specific topology, where trajectories that begin in close proximity diverge over time due to the system's sensitivity to initial conditions, eventually filling out the manifold.

An advantage of this dynamical systems perspective is that it enables the use of powerful mathematical tools to investigate the system's structure. One of these tools is \textbf{Takens' embedding theorem} \cite{takens2006detecting}, which states that the structure of a high-dimensional nonlinear dynamical system can be reconstructed from time series measurements of any single variable in the system---a process known as \textbf{state space reconstruction}. Building on this idea, Sugihara et al. \cite{sugihara2012detecting} proposed \textbf{Convergent Cross Mapping} (CCM), a method for detecting causal interactions in deterministic chaotic systems. These methods have been originally applied to ecological dynamics with great success and have since been used to recover directed connections in various neural data modalities, including neuronal cultures \cite{tajima2017locally}, EcoG data \cite{tajima2015untangling}, and crucially, spiking activity from the monkey brain \cite{nejatbakhsh2023predicting}. In the latter case, a CCM-based method successfully identified \textbf{hub electrodes} from spontaneous activity; stimulation of these hubs produced significantly larger effects on network activity than stimulation of non-hub sites. Notably, traditional methods, such as TE and linear GC, were inferior, failing to predict the magnitude of stimulation effects.

To date, electrical stimulations have been limited to one or two electrodes at a time. Once hub electrodes are identified, can we design more complex, multi-electrode stimulation patterns to steer neural activity in the recorded area toward a desired state? Answering this question has far-reaching implications for the development of clinical interventions. 

\subsection{Delay embedding of neural activity in the nonlinear regime}

Overwhelming evidence suggests that brain activity during resting epochs exhibits features typical of nonlinear dynamics, such as multistability and switching dynamics, at both the level of local circuits \cite{kenet2003spontaneously,mazzucato2015dynamics} and the whole brain \cite{deco2012ongoing}. Spontaneous neural activity is high dimensional \cite{manley2024simultaneous}, and a potential explanation is that it traces out a high-dimensional chaotic attractor. Fortunately, such complex systems are amenable to data-driven analysis using tools based on the concept of \textbf{delay embedding}. 

Although dynamics are highly nonlinear in the original activity space, they become approximately linear in a high-dimensional auxiliary \textbf{delay space}. In this space, the system's state at the next timestep can be predicted as a linear combination of its current high-dimensional state \cite{arbabi2017ergodic}. This formalism---known as the \textbf{Koopman operator} framework---is discussed in Section \ref{Koopman}. This perspective brings us one step closer to our final goal: designing a control signal to steer the system toward a desired final state. Once we have a linear dynamical system, we can design an optimal control signal that minimizes a user-defined cost function to drive the system from an initial state to a target state. Designing such a control signal becomes easy thanks to the linear dynamics in the delay space. This method known as \textbf{Hankel DMD with control} is discussed in section \ref{control}.

\section{Takens' theorem and delay embedding}\label{Taken's theorem}

\begin{figure}[h]
    \centering
    \includegraphics[width=0.8\linewidth]{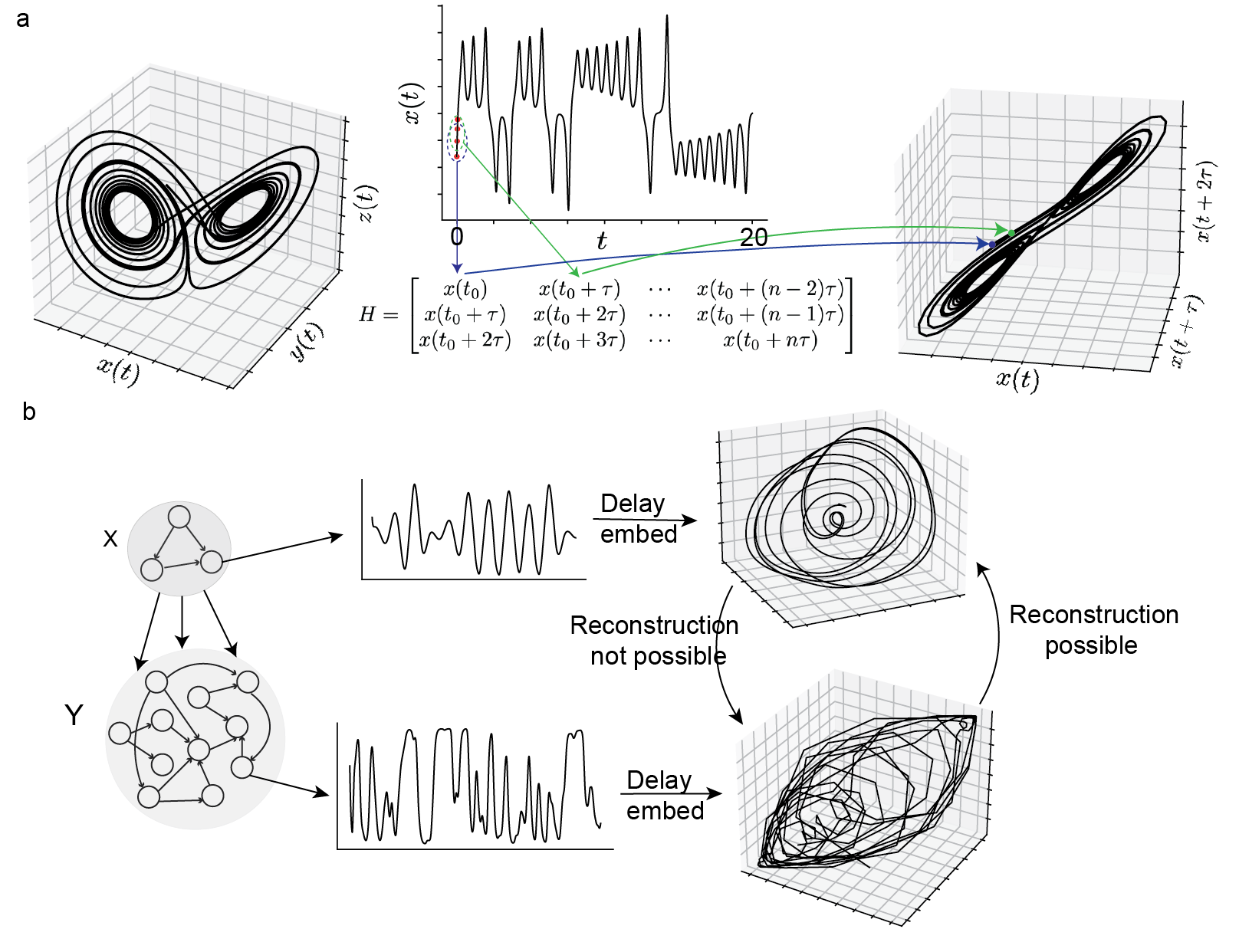}
    \caption{\textbf{a)} Left: Lorenz attractor in dynamical state space. Middle: $x(t)$ and 3D Hankel matrix constructed from $X(t)$. Right: Trajectory traced out in the delay space by the points in the Hankel matrix. \textbf{b)} Left: Schematic of a network with 2 subnetworks X and Y. X consists of 3 neurons that support a chaotic trajectory of a combinatorial threshold linear network (CTLN) \cite{parmelee2022core}. Subnetwork $X$ has unidirectional projections to subnetwork $Y$, consisting of $n=100$ neurons and dynamical equation $\dot{y}= W \, tanh(y), \; W \sim \mathcal{N}(0,g^2/n)$, with $g=1$. Middle: Time-series of a neuron $x_1$ in X (top) and a neuron $y_1$ in Y (bottom). Right: 3-dimensional delay embedding of the timeseries of $x_1$ (top) and $y_1$ (bottom). The delay embedded trajectory of $x_1$ can be reconstructed from the delay embedded trajectory of  $y_1$, but not vice versa, which defines a directed causal connection $x_1 \rightarrow y_1$ }
    \label{fig:lorenz_sys_delay_embed}
\end{figure}

We demonstrate \textbf{state space reconstruction} via Takens' theorem using the Lorenz system as an example, as shown in Figure \ref{fig:lorenz_sys_delay_embed}. In panel a, we show the chaotic attractor generated by the Lorenz system's trajectory across its 3 variables:  $x(t)$, $y(t)$, $z(t)$. Suppose that we can measure only one of these variables, $x(t)$. We can create a $D$ dimensional delay embedding of $x(t)$, where each point in the delay space is given by $[x(t), x(t+\tau), \cdots, x\left(t+(D-1)\tau \right)]$, sampling $x(t)$ at steps of $\tau$. Figure \ref{fig:lorenz_sys_delay_embed} shows a 3 dimensional delay embedding of $x(t)$. The matrix formed by defining each delay-embedded point as a column vector is known as the \textbf{Hankel matrix}, $H$ (Fig. \ref{fig:lorenz_sys_delay_embed}a). 

Takens' theorem states that the full attractor in the $[x,y,z]$ space  can be continuously deformed to get to the attractor on the $x$-only delay space and vice versa. In other words, the trajectory reconstructed from $x(t)$ alone preserves the topological structure of the original state space. Such mappings from one manifold to another are known as \textbf{diffeomorphisms}---they preserve the topological properties such as the two-lobed structure of the Lorenz attractor in our example.

\section{Functional Causal Flow using Delay Embedding}  \label{FCF}
Can we use state space reconstruction to map out the \textbf{causality structure} within a population of neurons? If we can infer the causal links between all pairs of observed neurons, we can then identify the \textbf{hubs}: the neurons with the largest causal efficacy which become the targets of our perturbation experiments. 

To illustrate this approach, we consider a recurrent neural network consisting of two groups of neurons $X=[x_1(t), \cdots, x_n(t)]$ and $Y=[y_1(t), \cdots, y_n(t)]$, where neurons in $X$ directly project to neurons in $Y$ but not vice versa. Both subnetworks exhibit time-varying activity, generating a chaotic attractor (Fig. \ref{fig:lorenz_sys_delay_embed}b). Since neurons $y_i(t)$ are downstream of neurons $x_j(t)$ (i.e., the activity of $y_i$ depends on $x_j$  but not vice versa), it follows that $x_j(t)$ can be reconstructed from the delay embedding of $y_i(t)$ for any $i \in [1, n], j \in [1, m]$. In contrast, $y_i$ cannot be reconstructed from $x_j$. This \textbf{asymmetry in reconstructability} reveals the causal directionality. This is the core idea behind the computation of \textbf{Causal Flow}, as shown in Figure \ref{fig:lorenz_sys_delay_embed}b. The causal flow between $y_i$ and $x_j$ is defined by how well $x_j$ can be predicted from past $y_i$. The exact measure, methods for establishing statistical significance, and procedure for optimizing delay dimension $D$ and time step $\tau$ are explained in \cite{nejatbakhsh2023predicting}. Crucially, causal flow predicts the effects of direct perturbation of neural activity. Thus, as a first step toward designing control signals that steer networks to particular dynamical states, the hub nodes in the network can be identified from resting-state data using this method. Restricting stimulations to the hub nodes is expected to produce the strongest perturbation effects.

\begin{figure}[h]
    \centering
    \includegraphics[width=\linewidth]{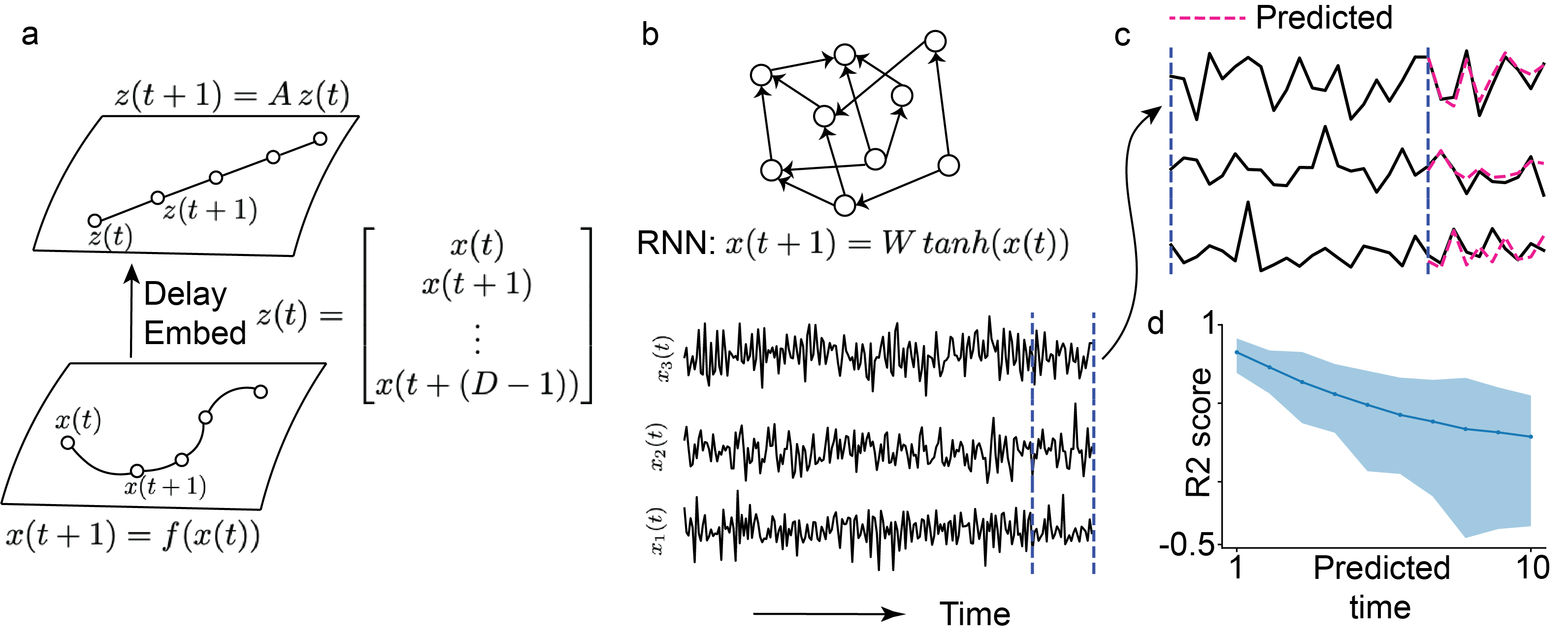}
    \caption{\textbf{a)} Schematic of Koopman operator. Nonlinear dynamics of a system becomes linear in higher dimensional delay space. \textbf{b)} Schematic of an RNN (top) and timeseries of 3 representative neurons (bottom). \textbf{c)} In delay space, the Koopman operator can be approximated by the matrix which maps one column of the Hankel matrix to its adjacent column on the right. This matrix $A$ can be obtained by solving the linear least squares problem $H_1= A\, H_0$ where $H_0$ consists of the first $n-1$ columns of the Hankel matrix $H$, and $H_1$ consists of the last $n-1$ columns of $H$. We reduce back to the $x$ activity space via $x=Cz$ where $C= [ 0 \; I_{n\times n}]$, where $n$ is the length of $x$. Predicted future trajectories match the simulated trajectories quite well. \textbf{d)} $R^2$-score of the predicted trajectory for the subnetwork Y in \ref{fig:lorenz_sys_delay_embed} across 10 predicted timesteps demonstrates high prediction accuracy for short time horizons, but a rapid decline of accuracy for longer horizons. The line represents the median across 100 initial conditions, and the shaded region indicates the 95\% confidence interval. }
    \label{fig:dmd_koopman}
\end{figure}

\section {Koopman operator and delay embedding}\label{Koopman}
Once the hub neurons are identified, the next step is to design a stimulation protocol capable of driving neural activity toward the desired outcome, such as evoking a specific population activity pattern to influence behavior. To leverage the powerful tools of linear control theory for this problem, we must first linearize the highly nonlinear dynamics observed in biological neural populations.

Gaining analytical insight into nonlinear dynamical systems---and designing effective control signals for them---remains a significant challenge. Fortunately, delay embedding provides a pathway to overcome this obstacle. By projecting the nonlinear dynamics of a chaotic system into a high-dimensional delay space, it becomes possible to approximate the system as linear~\cite{arbabi2017ergodic}. Figure \ref{fig:dmd_koopman}a shows this concept schematically. The linear operator that advances a point in the delay space trajectory from one time step to the next is known as the Koopman operator. In Figure \ref{fig:dmd_koopman}, we apply this method to predict future time steps in the trajectory of a chaotic recurrent neural network (RNN).

For large networks, it is possible to reduce computational costs and complexity of experimental measurements by delay-embedding only a small subset of nodes. This is often sufficient because Takens' theorem suggests that the full dynamics of the system can be reconstructed from partial measurements. Once we linearize the dynamics with the Koopman operator, we can apply linear control techniques in the delay space to design optimal control signals---an otherwise very hard or intractable problem for nonlinear systems \cite{korda2018linear}.

\begin{figure}[h]
    \centering
    \includegraphics[width=\linewidth]{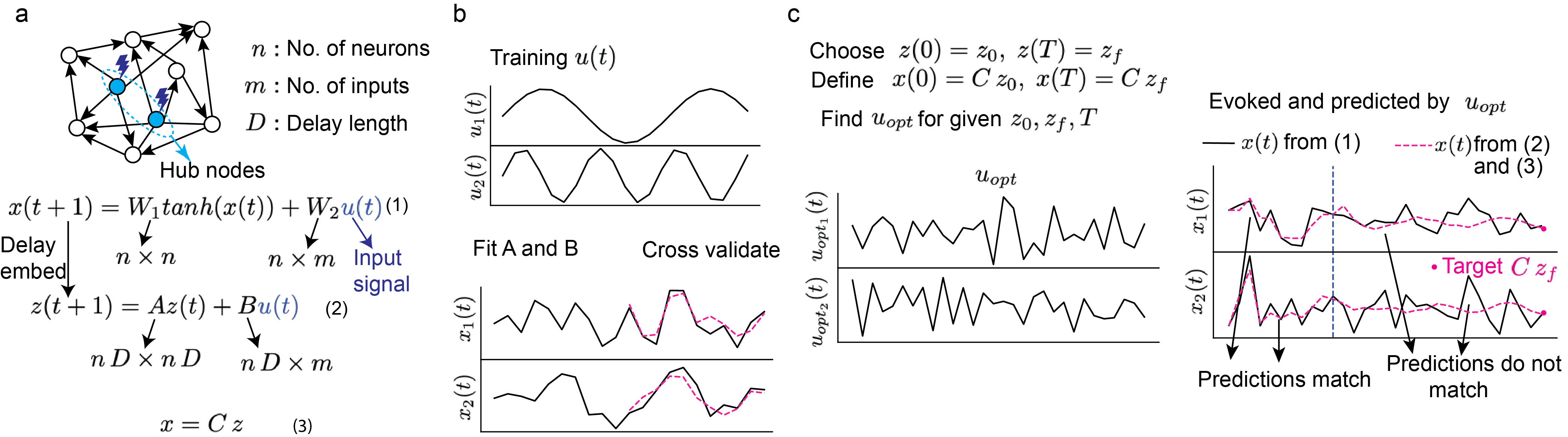}
    \caption{ \textbf{a)} Schematic of an RNN with hub nodes inferred via Causal Flow shown in light blue, where current $u(t)$ is injected during a training period used to fit matrices $A$ and $B$ to observed trajectories. \textbf{b)} Examples of training signals used in the simulations. Trajectories of two neurons in the RNN with the predicted trajectories from the fitted $A$ and $B$ for the last 10 timesteps. \textbf{c)} The minimum energy control input $u_{opt}(t)$ (two components, left) is estimated to drive the delay-space trajectory from $z_0$ to $z_f$ and the corresponding trajectory from $x_0=C \,z_0$ to $x_f=C \, z_f$ in the activity space (right). 
    %2 components of $u_{opt}(t)$ (top). 
    In the evoked activity plots (right), black lines show simulated trajectories from Eq(1) and pink dashed line shows predicted trajectories from Eq(2) and Eq(3) in panel a.}
    \label{fig:delay_dmd_with_control}
\end{figure}

\section{Control of nonlinear system using Koopman formalism}\label{control}
After identifying the hubs for stimulation and linearizing the nonlinear dynamics using the Koopman operator, the stage is set for designing a control signal to drive neural activity (or behavior) toward a desired target state (Figure \ref{fig:delay_dmd_with_control}a.) We can generalize this procedure to the case of dynamical systems driven by an external input: $x(t+1)=g(x(t),u(t))$, where $u(t)$ is the input.

As before, we delay-embed $x(t)$, but  instead of fitting a single matrix $A$ to the delay vector $z(t)$, we now fit two matrices, $A$ and $B$, to capture both the intrinsic dynamics and the effect of the input 
\begin{equation}\label{z_dyn}
z(t+1)=Az(t)+Bu(t)
\end{equation}
This algorithm, known as \textbf{dynamic mode decomposition (DMD) with control} \cite{proctor2016dynamic}, allows us to estimate $A$ and $B$ from a training period in which currents $u_i(t)$ are injected into the network hub nodes (previously identified from spontaneous activity using Causal Flow). This workflow is shown for our RNN example in Figure \ref{fig:delay_dmd_with_control}b,c. 

Our goal is to  drive the system from an initial state $z(0)=z_0$ to a target state $z(T)= z_f$ in delay space, which correspond to $x(0)=C \, z_0$ to $x(T)=C z_f$ in the original neural activity space. DMD with control achieves this goal by determining the optimal $u$, subject to two sets of constraints: the dynamical equations in delay space, involving matrices $A$ and $B$; and a control energy minimization criterion: $E=\sum_{t=0}^{t=T} |u(t)|^2$, where $E$ quantifies the total stimulation ``energy'' and $T$ is the control horizon (i.e., number of stimulation steps). $E$ can be interpreted as the total injected current into the circuit, and its minimization ensures that stimulation remains efficient and biologically safe for neural tissue. In Figure \ref{fig:delay_dmd_with_control}c, we derived the minimum energy control input , where $A$ and $B$ were estimated from a training period \cite{proctor2016dynamic}. We find that the evoked $\hat{x}_f$ closely approximates the desired target $x_f$, even though agreement between the predicted and evoked trajectories gradually deteriorates over longer horizons.

A key challenges in this approach is balancing the tradeoff between control horizon and prediction horizon. On one hand, shorter control horizon or stimulation timesteps result in large control energy and failure to reach the target state \cite{pasqualetti2014controllability}. On the other hand predicted trajectories match the evoked trajectories only for short periods of time (Fig. \ref{fig:dmd_koopman}d, Fig. \ref{fig:delay_dmd_with_control}d). Thus, a right balance must be found between prediction and control to drive the nonlinear system to the desired target $x_f$ using control signals $u$ that minimize user-defined cost functions and satisfy experimental constraints. We found that even though, the predicted and evoked trajectory do not match beyond the prediction horizon, they converge at the desired target state (Figure \ref{fig:delay_dmd_with_control}d), thus demonstrating the feasibility of the method outlined above in our RNN example.

\section{Conclusion}

Recent advances in neurotechnology have made it possible to deliver increasingly complex spatio-temporal stimulation patterns. However, the number of possible stimulation protocols grows exponentially with the number of neurons and timesteps, making the design space prohibitively large. To navigate this space effectively, three key components are required. First, designing a clearly defined objective for the desired effect of the stimulation, for example, as discussed in this review, to drive neural activity toward a target state via control theory. Second, identifying a small number of high efficacy targets for stimulation, such as the hub neurons we inferred from spontaneous activity via Causal Flow. Third, developing accurate computational models capable of predicting nonlinear neural dynamics during perturbations, such as using the Koopman operator.

Our proposed workflow combines hub identification via Causal Flow, neural dynamical modeling via Koopman theory, and linear control for achieving a target pattern or behavior. In particular, Koopman theory provides a powerful framework for analyzing nonlinear systems by lifting their dynamics into a higher-dimensional linear space. While this projection ideally requires an infinite-dimensional space to exactly capture the true operator, a finite-dimensional approximation—obtained via linear regression in equation \eqref{z_dyn}—can still predict system dynamics accurately over short time horizons. This is often sufficient for control objectives that operate on short timescales. Moreover, the linear structure enables the use of established tools from linear control theory, which offer strong theoretical guarantees and computationally efficient optimization \cite{korda2018linear}. 

Careful experimental design and execution will be needed to test whether the theoretical approach outlined above bears out in the empirical domain. For the first step, recent work showed that hub identification via Causal Flow accurately predicts stimulation effects in primate cortex \cite{nejatbakhsh2023predicting}. For the second step, Koopman theory has been shown to predict spontaneous movement time series in worms \cite{ahamed2021capturing} and identify sleep spindle network from Ecog data \cite{brunton2016extracting}, however, it has not been applied to spiking neural data yet. Likewise, linear control in delay space using the DMD with control algorithm has not been applied to brain-computer interfaces. Developing such new experimental paradigms remains a central challenge for neuroscience and an important direction for future research.

\section*{Acknowledgements}
L.M. was partially supported by National Institutes of Health
Awards R01NS118461, R01MH127375 and R01DA055439 and National Science Foundation CAREER Award 2238247. R.K. was supported by the Simons Collaboration on the Global Brain (542997 and 988247),
McKnight Scholar Award, Pew Scholarship in the Biomedical Sciences, and National Institute
of Mental Health (R01 MH109180 and R01 MH127375). 

\section*{Declaration of interest}
None

\section*{Annotated Bibliography}

\noindent\cite{sugihara2012detecting}: **This paper introduces Convergent Cross Mapping (CCM), a method which relies on nonlinear state space reconstruction of one variable from another variable in a nonlinear dynamical system to establish causal connection between these variables.

\noindent\cite{nejatbakhsh2023predicting}: **This paper applies CCM to spiking activity recorded from macaque cortex to find hub electrodes defined by maximum number of outgoing connections. The authors find that electrically stimulating hub electrodes create a greater shift in spontaneous activity than stimulating non-hub electrodes.

\noindent\cite{arbabi2017ergodic}: **This paper shows that dynamics of a nonlinear chactic system becomes linear when projected to a high dimensional delay space and the matrix which evolves the delay vectors can be approximated by linear regression.

\noindent\cite{korda2018linear}: **This paper extends linearizing dynamics in high dimensional delay spaces to nonlinear controlled system to express it as a linear control system. They show that designing controllers for the high dimensional linear system is computationally easier than the nonlinear system.

\noindent\cite{proctor2016dynamic}: **In this paper the authors develop a method to to disambiguate between the underlying dynamics and the effects of external signals or control inputs for a dynamical system with inputs from the timeseries data of the system and inputs.

\bibliographystyle{naturemag}
\bibliography{\jobname}

\end{document}